\documentclass[journal]{IEEEtran}
\usepackage{cite}

\ifCLASSINFOpdf
  \usepackage[pdftex]{graphicx}
	\usepackage[table]{xcolor}
  \graphicspath{{BeamwidthOptimization_images/}}
  \DeclareGraphicsExtensions{.jpg,.eps}
\else
\fi
\usepackage{amsmath}
\usepackage{array}

\ifCLASSOPTIONcompsoc
  \usepackage[caption=false,font=normalsize,labelfont=sf,textfont=sf]{subfig}
\else
  \usepackage[caption=false,font=footnotesize]{subfig}
\fi
\usepackage{url}

\usepackage[normalem]{ulem}

\begin{document}
%
\title{Phase-Only Beam Broadening of Contiguous Uniform Subarrayed Arrays Utilizing Three Metaheuristic Global Optimization Techniques}
%
%
%

\author{Barry~Daniel,~\IEEEmembership{Senior Member,~IEEE}, Carl Edwards, and Adam~Anderson,~\IEEEmembership{Senior Member,~IEEE}}

%
%

\markboth{arXiv.org}%
{Daniel \MakeLowercase{\textit{et al.}}: Journal Paper}
%



\maketitle

\begin{abstract}
    Radar beam broadening provides continuous coverage of a wider angular extent. 
    While many methods have been published that address beam broadening of traditional (non-subarrayed) arrays, there is a knowledge gap in the published literature with respect to efficient and effective beam broadening of contiguous uniform subarrayed arrays.
    This paper presents efficient and effective methods for beam broadening of contiguous uniform subarrayed arrays where elements of the array are grouped together to have the same element excitations. Particularly, this paper focuses on phase-only optimization to preserve maximum power output. 
    The high dimensionality of the solution space of possible phase settings causes brute force techniques to be infeasible for exhaustively evaluating the entire space. This paper presents three metaheuristic global optimization techniques that efficiently and effectively search for optimal phase values in this large solution space that satisfy the desired broadened pattern. The techniques presented in this paper are simulated annealing, genetic algorithm with elitism, and particle swarm optimization. These techniques are evaluated on idealized 40x40 and 80x80 element rectangular arrays with 5x5 element subarrays.
    The results of this study show that as configured in this paper the simulated annealing and particle swarm techniques outshine the genetic algorithm technique for 40x40 and 80x80 rectangular arrays grouped into contiguous uniform 5x5 element subarrays.
\end{abstract}


%
\IEEEpeerreviewmaketitle

\section{Introduction}
    \IEEEPARstart{C}{ontinuous} coverage of wide areas is desirable in many application realms including radar and cellular communications.  Radars are challenged with finding efficient means to scan wide angular extents to provide detection, tracking, characterization and measurement of objects within their coverage area.  Similarly, the cellular communication industry needs to efficiently provide voice and data communications over very broad coverage areas \cite{Rajagopal}.  This coverage problem can be addressed by steering the main beam across the coverage area using mechanical or electronic means, but beam broadening has the advantage of continuous coverage of a wider angular extent instead of using a time-multiplexing scheme as utilized by beam steering.
      
    In a phased array radar, broadening of the main beam can be accomplished by either adjusting the amplitude or phase of the individual elements of the array or by adjusting both the amplitude and phase. In today's modern phased arrays that utilize solid-state transmit amplifiers which operate most efficiently in saturation, it is often desired to utilize phase-only approaches to beam broadening as presented in \cite{Rajagopal,Kerce-Brown-Mitchell,Huang-Liu-You-Yang} to optimize power efficiency.  However, these published methods of beam broadening assume that the phase of each element in the array can be independently adjusted which is not true of contiguous uniform subarrayed arrays.
    
    In a contiguous uniform subarrayed array architecture the elements are grouped together in a manner that only allows the amplitude and phase of each subarray, not each element, to be adjusted. Contiguous uniform subarrayed arrays are commonly used in large array designs to reduce cost by limiting the numbers of control elements.  However, the subarrayed architecture also makes it more difficult to form a low-ripple broadened main beam and to reduce the sidelobes.  There is currently a gap in the available literature regarding the efficiency of global optimization methods to synthesize antenna patterns with a broadened main beam for subarrayed architectures when the synthesis is constrained to be phase-only, utilize contiguous uniform subarrays, and produce low sidelobes. \cite{Lee-Tseng,Rocca-Manica-Massa,Rocca-Manica-Azaro-Massa} address synthesis with subarrayed architectures but they concentrate on optimization of combined sum and difference patterns, not broadening the main beam. \cite{Haupt} proposes a synthesis method for subarrayed arrays but this method is amplitude only, utilizes subarrays of unequal sizes, and does not address beam broadening. \cite{Rocca-Haupt-Massa,Manica-Rocca-Massa,Oliveri,Manica-Rocca-Oliveri-Massa} approach our desired domain space but utilize hybrid subarrays where the amplitude is dependent on the subarray amplitude but the phase of each element in the subarray can still be adjusted independently. The closest published material that this author has found is \cite{Rajagopal} which presents a beam broadening approach for subarrayed architectures where the synthesis is constrained to be phase-only and utilize contiguous uniform subarrays, but this method constrains the widened full array beam widths to only those beam widths that are multiples of the subarray beam width. This method also produces unacceptable sidelobe levels.  
    
    Metaheuristic global optimization methods such as Simulated Annealing (SA) \cite{trastoy2001phase}, Genetic Algorithms (GA) \cite{wen2016wide, bhargav2013multiobjective}, and Particle Swarm Optimization (PSO) \cite{yang2014adjustment, song2010application} have been used to find solutions for several radar array architectures utilzing both amplitude or phase adjustment. These optimization problems cannot be optimized efficiently utilizing brute force techniques due to computational limits.  These global optimization algorithms are capable of searching complicated solution spaces with many local optima to find the global optimum. Due to the dimensionality and computational difficulty of radar beam broadening problems, these techniques are well-suited for synthesis of antenna patterns using various cost evaluation functions. 
    
    This paper compares three of the most popular global optimization algorithms and their ability to generate broadened main beam patterns in subarrayed radar architectures. It compares the efficiency, the rate at which a sufficient solution is found, and effectiveness, the relative quality of the solutions, of these techniques in terms of cost function evaluation of the pattern generated from a given phase setting for 40x40 element arrays and 80x80 element arrays. Simulation results of pattern optimization for these subarrayed arrays are given to demonstrate the validity of these techniques and their comparability. 
    
    The mathematical model of the optimization algorithms utilized in this paper are presented in Section \ref{Mathematical Model Section} which is followed by Section \ref{Cost Function Section} which presents the cost function used.  Section \ref{Results Section} shows the simulated results of the various global optimization algorithms. Finally, Section \ref{Conclusions Section} presents the conclusions that are drawn from the results.

\section{Mathematical Model - Global Optimization Algorithms}
\label{Mathematical Model Section}
    \subsection{Simulated Annealing} 
        Simulated annealing (SA) is inspired by the annealing process in metallurgy where material is heated and cooled to increase the size of its crystals and reduce their defects.  
        This optimization technique can be utilized to find an approximation of a global minimum for functions with a large number of variables and a large solution space. 
        The notion of slow cooling implemented in the simulated annealing algorithm is interpreted as a slow decrease in the probability of accepting worse solutions as the solution space is explored. Accepting worse solutions is a fundamental property of metaheuristics because it allows for a more extensive search for the global optimal solution. In general, the simulated annealing algorithms work as follows. At each iteration, the algorithm randomly selects a solution close to the current one, measures its quality, and then decides to move to it or to stay with the current solution based on either one of two probabilities between which it chooses based on the relative fitness of the new solution to the old one. During the search, the temperature is progressively decreased from an initial positive value towards zero and affects the probability of changing solutions: at each temperature change, the probability of moving to a better new solution is true, but the probability of moving to a worse new solution is progressively changed towards zero.
            
        The SA process used in this paper is derived from the implementation described in \cite[Chapter~7]{skiena1998algorithm}. The variables utilized are defined as:
    
        \begin{itemize}
            \item $S$: The candidate solution of element excitations.
            \item $T$: The current annealing temperature. This starts at $T_0 = 1$.
            \item $T_{stops}$ The number of temperature changes desired.  
            \item $\alpha$: The exponential temperature decrease constant.
            \item $k$: Normalize the fitness function so high temperatures are more reactive than low temperatures.
            \item $C$: A function that assigns an evaluation score to a given candidate set of element excitations.
            \item $i$: The number of iterations of the algorithm between temperature change.
            \item $d$: Number of variables (dimensions of solution space).
        \end{itemize}
        
    
        The implemented SA process can be described as:
    
        \begin{enumerate}
            \item Randomly generate a candidate $S$ of size $N$ element excitations.
            \item Synthesize the array pattern and evaluate the generated pattern against the desired pattern via a cost function, $C$.
            \item For $T_{stop}$ temperature changes.
            \begin{enumerate}
                \item For $i$ iterations 
                \begin{enumerate}
                    \item Generate a neighboring solution by randomly changing one value in $S$.
                    \item Synthesize the array pattern and calculate the cost.
                    \item If the new solution's cost $C_{new}$ is less than the old solution's cost $C_{old}$, replace $S$ with it. If it is not, replace $S$ with the new solution if
                        $$\text{rand()} < e^{\frac{C_{old}-C_{new}}{kT}}$$ 
                \end{enumerate}
                \item $T \leftarrow \alpha T$
            \end{enumerate}
            \item Return the solution $S$.
        \end{enumerate}
        
    \subsection{Genetic Algorithm} 
        A genetic algorithm (GA) is inspired by the process of natural selection and relies on bio-inspired operators such as mutation, crossover and selection to evolve an initial population of candidate solutions toward better solutions. Each candidate solution has a set of properties (its chromosomes or genotype) which can be mutated and altered. The evolution usually starts from a population of randomly generated individuals, and is an iterative process, with the population in each iteration called a generation. In each generation, the fitness of every individual in the population is evaluated; the fitness is usually the value of the objective function in the optimization problem being solved. The more fit individuals are stochastically selected from the current population, and each individual's genome is modified (recombined and possibly randomly mutated) to form a new generation. In addition, elite members of the population with the highest fitness are preserved. The new generation of candidate solutions is then used in the next iteration of the algorithm. The algorithm terminates when either a maximum number of generations has been produced, or a satisfactory fitness level has been reached for the population.
        
        For our beam broadening array pattern optimization problem the population includes candidate sets of element excitations for the array. Therefore, in genetic algorithm terminology, a set of element excitations for each element in the array is the chromosome and each individual element citation is the gene.  The new population of sets of element excitations is generated as:
    
        The GA process used in this paper is derived from the implementation described in \cite{RN404} but minimizes a cost function instead of maximizing a fitness function. In addition, elite members of the population are introduced due to analysis in \cite{rudolph1994convergence}. The variables utilized in the description of the GA process are defined as:
    
        \begin{itemize}
            \item $P$: The population of candidate sets of element excitations.
            \item $P_{new}$: The new population of candidate sets of element excitations derived from the previous population.
            \item $C$: A function that assigns an evaluation score to a given candidate set of element excitations.
            \item $G$: The maximum number of generations.
            \item $N$: The number of candidate sets of element excitations to be included in the population.
            \item $r$: The fraction of the population to be replaced via the crossover technique at each step.
            \item $m$: The mutation rate.
        \end{itemize}
    
        The implemented GA process can be described as:
    
        \begin{enumerate}
            \item Randomly generate a population, $P$, of size $N$ candidate sets of subarray excitations.
            \item For each candidate set of subarray excitations, synthesize the array pattern and evaluate the generated pattern against the desired pattern via a cost function, $C$.
            \item For $G$ generations of solutions
            \begin{enumerate}
                \item Create a set of parents by probabilistically selecting $rN$ members of $P$. The probability $Pr(n)$ of selecting candidate set $n_{i}$ from $P$ is given by 
                \begin{equation}
                  Pr(n_{i}) = \frac{\max(C(n)) - C(n_{i})}{\sum_{j=1}^{N}(\max(C(n)-C(n_{j}))}
                \end{equation}
                if all the values of $C(n_{i})$ are the same, $Pr(n_{i}) = \frac{1}{N}$.
                \item For each pair of candidate sets from $P$, $\langle n_1, n_2\rangle$, produce two offspring by applying the crossover operator.  Add all offspring to $P_{new}$.
                \item For each of the members of $P_{new}$, perform the following mutation operator from \cite{rudolph1994convergence}: for each gene, if $\text{rand()} < m$, mutate it. 
                \item Add the $(1-r)N$ members of $P$ with the lowest cost into $P_{new}$ as elites.
                \item Update the population: $P \leftarrow P_{new}$.
                \item For each candidate set of subarray excitations, synthesize the array pattern and evaluate the generated pattern against the desired pattern via a cost function, $C$.
            \end{enumerate}
            \item Return the candidate set from $P$ that has the lowest cost function value.
        \end{enumerate}
    
    \subsection{Particle Swarm Optimization}
        Particle swarm optimization (PSO) functions by following a physical analogy of particles swarming around a solution with a given velocity. Particles solutions are randomly created and move around the solution space with a velocity created from their personal best solution, the global best solution, and inertia. This allows the particles to explore the solution space while still being influenced by the swarm intelligence. 
        Due to the cyclical nature of the solution space, $V_{max}$ was not found to be necessary because of angle wraparound of phase degree values, since it would have been set to the dynamic range of each variable which is $2\pi$ radians.

        The PSO used in this paper is derived from the implementation described in \cite{shi2001particle}. With regard to the synthesis of array patterns, a particle is defined as a set of subarray excitations. The variables utilized in the description of the PSO are defined as:
    
        \begin{itemize}
            \item $x$: The particles (subarray excitations) in the population.
            \item $v$: The velocities of the particles in the population.
            \item $p$: The best solution found by a particle.
            \item $c_1$ and $c_2$: Acceleration constants for the velocity update rule. 
            \item $\omega$: The inertia of the particle: used to perserve some of the prior velocity. 
            \item $C$: A function that assigns an evaluation score to each particle.
            \item $N$: The number of particles to be included in the population of particles.
            \item $i$: An index of which particle is being evaluated.
            \item $g$: The index of the global best particle.
            \item $d$: Number of variables (dimensions of the solution space) or independent subarray excitations.
        \end{itemize}
    
        The implemented PSO process can be described as:
    
        \begin{enumerate}
            \item Randomly generate a population of $N$ where each particle is defined by a position, $x_i$, and a velocity, $v_i$. Generate velocities for each particle in the population from $[-2\pi, 2\pi]$ phase values. 
            \item While the number of iterations is less than some given maximum number, repeat the following steps.
            \begin{enumerate}
                \item For each particle, synthesize the array pattern and evaluate the generated pattern against the desired pattern via a cost function, $C$.
                \item Check each particle if there is a new personal best $p_{id}$ and check for a new global best $p_{gd}$.
            
                \item Calculate the velocity given by 
                \begin{multline}
                    v_{id} = \omega*v_{id} + c_1*\text{rand()}*(p_{id}-x_{id}) + \\c_2*\text{rand()}*(p_{gd}-x_{id})
                \end{multline}
                \label{velocity equation}
                \item Add the velocity to the existing position to move the particles to new locations in the solution space.
                $$ x_{id}=x_{id}+v_{id} $$
            \end{enumerate}
            \item Return the candidate set from $P$ that has the lowest cost function value.
        \end{enumerate}

        For our purposes the velocity calculation in Step \ref{velocity equation} was modified from the equation given in \cite{shi2001particle} because phase is a circular function and wraps around at $2\pi$ or 360 degrees. For example, if the current particle, $x_{id}$, is at 10 degrees and the local best particle, $p_{id}$, is at 290 degrees the shortest distance between the particles is not $290 - 10 = 280$ degrees. The shortest distance is actually through the wrap-around at 360 degrees. If the calculated local distance or global distance was greater than 180 degrees, then 360 degrees was subtracted from the calculated distance.  Similarly, if the calculated local distance or global distance was less than -180 degrees, then 360 degrees was added to the calculated distance.  This adjustment ensures that the velocity of the particle is at the correct amplitude and direction to more quickly converge to a global solution.
        
\section{Mathematical Model - Cost Function}
\label{Cost Function Section}
    In this paper a cost function determines how close the generated array factor, $AF_{\mathrm{gen}}$, is to the desired array factor, $AF_{\mathrm{des}}$. The desired array factor is defined as a normalized flat-top function with a main beam and sidelobes.  The main beam consists of a desired angular width with an amplitude of 0 dB.  The sidelobes are the points in $AF_{\mathrm{des}}$ outside this main beam with a desired sidelobe level defined in negative dB.  $AF_{\mathrm{gen}}$ is computed by taking the IFFT of the element excitations and is normalized before comparing it to $AF_{\mathrm{des}}$. The cost function, $C$, is defined as:
    \begin{equation}
        C = \log({E}_{\mathrm{mb}} + {E}_{\mathrm{sl}})
        \label{Cost Function}
    \end{equation}
    where ${E}_{\mathrm{mb}}$ is the error calculated for the points that are in the main beam and below 0 dB, $\theta_{\mathrm{mb}}$, and is defined as:
    \begin{equation}
        {E}_{\mathrm{mb}} = \sum_{\theta_{\mathrm{mb}}} (\lvert AF_{\mathrm{des}}(\theta_{\mathrm{mb}}) \rvert - \lvert AF_{\mathrm{gen}}(\theta_{\mathrm{mb}}) \rvert)^2 
    \end{equation}
    and ${E}_{\mathrm{sl}}$ is the error calculated for the points that are outside the main beam and above the desired sidelobe level, $\theta_{\mathrm{sl}}$, and is defined as:
    \begin{equation}
        {E}_{\mathrm{sl}} = \sum_{\theta_{\mathrm{sl}}} (\lvert AF_{\mathrm{gen}}(\theta_{\mathrm{sl}}) \rvert - \lvert AF_{\mathrm{des}}(\theta_{\mathrm{sl}}) \rvert)^2 
    \end{equation}.
        
    The logarithm is used in (\ref{Cost Function}) to reduce the values returned to a more condensed range so that the convergence rate of the cost function can be more easily visualized. For a more tangible performance measure, (\ref{Cost Function}) can be converted to a percentage pattern effectiveness metric as
    \begin{equation}
        P_{\mathrm{eff}} =  10^{-\sqrt{\exp({C})/({\beta \times 100)}}} \times 100,
        \label{Power Efficiency}
    \end{equation}
    where $\beta$ represents the number of angular points used to evaluate $AF_{\mathrm{gen}}$ vs. $AF_{\mathrm{des}}$.

\section{Results}
\label{Results Section}
    The results presented in this paper were generated using a computer simulation that synthesized an approximation of the array pattern where each element was assumed to be an isotropic radiator.  The array pattern  was generated with an FFT-based approach for computational efficiency.  This simulation contained adjustable parameters that allowed arbitrary arrays to be modeled based on the number of elements in each subarray, the number of subarrays in the array, the spacing between elements, and the number of bits for each phase shifter.  The desired antenna pattern was described using parameters for width (12 degrees) of the broadened main beam and maximum sidelobe level (-13 dB).  
    
    The simulation was configured to model  rectangular arrays of 40x40 and 80x80 elements where the elements were uniformly spaced a half wavelength apart.  The rectangular arrays were grouped into contiguous uniform subarrays of 5x5 elements; thus, the 40x40 array had 8x8 subarrays and the 80x80 had 16x16 subarrays.  Each array was also configured to utilize 6-bit phase shifters.  These array configurations create solution spaces with $(2^{6})^{64}$ and $(2^{6})^{256}$ possible permutations for each respective array.  
    
    A few techniques were utilized in the simulation to shrink these solution spaces. First, array symmetry is assumed in order to simplify the problem. Symmetry is created by setting subarrays equidistant from the array center equal to the same phase. Assuming symmetry reduces the 40x40 array to only needing 9 independent phase settings instead of 64, and the 80x80 array requires 32 instead of 256.  Also, since it is the relative phase of the elements and subarrays to each other that is important, the innermost subarrays were fixed to zero phase to reduce the solution space to 8 and 31 independent phase settings for each respective array. Therefore, assuming symmetry and fixing the center subarray phase to zero shrinks the solution spaces to $(2^{6})^{8}$ and $(2^{6})^{31}$ possible permutations for each respective array.
    
    In order to effectively evaluate the stochastic algorithms, we allowed each algorithm to converge until they had the opportunity to evaluate 20,000 solutions.   For each iteration of the algorithm, we recorded the best solution as convergence curves. Since the SA, GA, and PSO algorithms are not deterministic processes, they were executed 100 times with the same settings to obtain statistical significance. The convergence curves from each of the 100 executions were averaged to form a mean convergence curve for each algorithm. 
    
    For SA, constants in Table \ref{tab:SA_table} were selected from ranges in \cite{skiena1998algorithm}. $k$ was chosen in order to accept many solutions at the beginning but to accept fewer as the temperature neared 0. It was based on the cost obtained from $C$. The logarithm utilized in $C$ was important for ensuring that $k$ was a satisfactory value for the duration of the algorithm as the cost decreased. Since $C$ involves a 2D summation of errors, differing patterns can have magnitudes of difference in fitness, and the logarithm allows $k$ to better cover the range of costs. $\alpha$ was selected to generally cover the range of $T$ over the course of the algorithm: its value is fairly standard. In addition, the number of iterations between temperature changes was selected to allow an appropriate number of changes in the algorithm given the rough number of evaluations of $C$ desired. 
    \begin{table}[]
        \centering
      \renewcommand{\arraystretch}{1.2}
        \caption{Constants Used in SA}
        \begin{tabular}{|c c c c|}
            \hline
             $\alpha$ & $k$ & $i$ & $T_{stops}$\\
             \hline 
             0.97 & 0.01 & 200 & 100\\ 
             \hline
        \end{tabular}
        \vspace{.1 cm}
        \label{tab:SA_table}
    \end{table}
    
    Constants in Table \ref{tab:GA_table} were empirically determined using a limited parameter search and could likely be improved with a more extensive search. This more extensive search is left as a an area for future study.  Using a quarter of the population as elites in the parent selection was determined to improve convergence. Due to the use of elitism preserving genes in the GA, $r$ and $m$ were both set higher to increase the rate of testing new solutions. The selection of population size was based on the rule-of-thumb that population sizes between 20 - 50 are usually sufficient.  For the smaller 40x40 array with a smaller solution space, a population of size 25 was used. However, in the case of the 80x80 radar, due to the increased solution space size, a larger population size of 50 was used. 
    
    \begin{table}[]
        \centering
        \caption{Constants Used in GA}
      \renewcommand{\arraystretch}{1.2}
        \begin{tabular}{|c c c c|}
            \hline
             & $m$ & r & $N$ \\ 
             \hline 
             40x40 & 0.07 & 1 & 25 \\
             \hline
             80x80 & 0.07 & 1 & 50 \\
             \hline
        \end{tabular}
        \vspace{.1 cm}
        \label{tab:GA_table}
    \end{table}
    
    Constants in Table \ref{tab:PSO_table} utilized in the PSO algorithm were chosen based on \cite{shi2001particle}.  
    To allow a fair comparison between the GA and PSO algorithms, the same population sizes were used for each algorithm. 
    \begin{table}[]
        \centering
      \renewcommand{\arraystretch}{1.2}
        \caption{Constants Used in PSO}
        \begin{tabular}{|c c c c c|}
            \hline
             & $\omega$ & $c_1$ & $c_2$ & $N$ \\ 
             \hline 
             40x40 & 0.729 & 1.49445 & 1.49445 & 25 \\
             \hline
             80x80 & 0.729 & 1.49445 & 1.49445 & 50 \\
             \hline
        \end{tabular}
        \vspace{.1 cm}
        \label{tab:PSO_table}
    \end{table}
    
    \subsection{40x40 Results}
    
\begin{figure}[!ht]
            \centering
                \includegraphics[width=3.2 in]{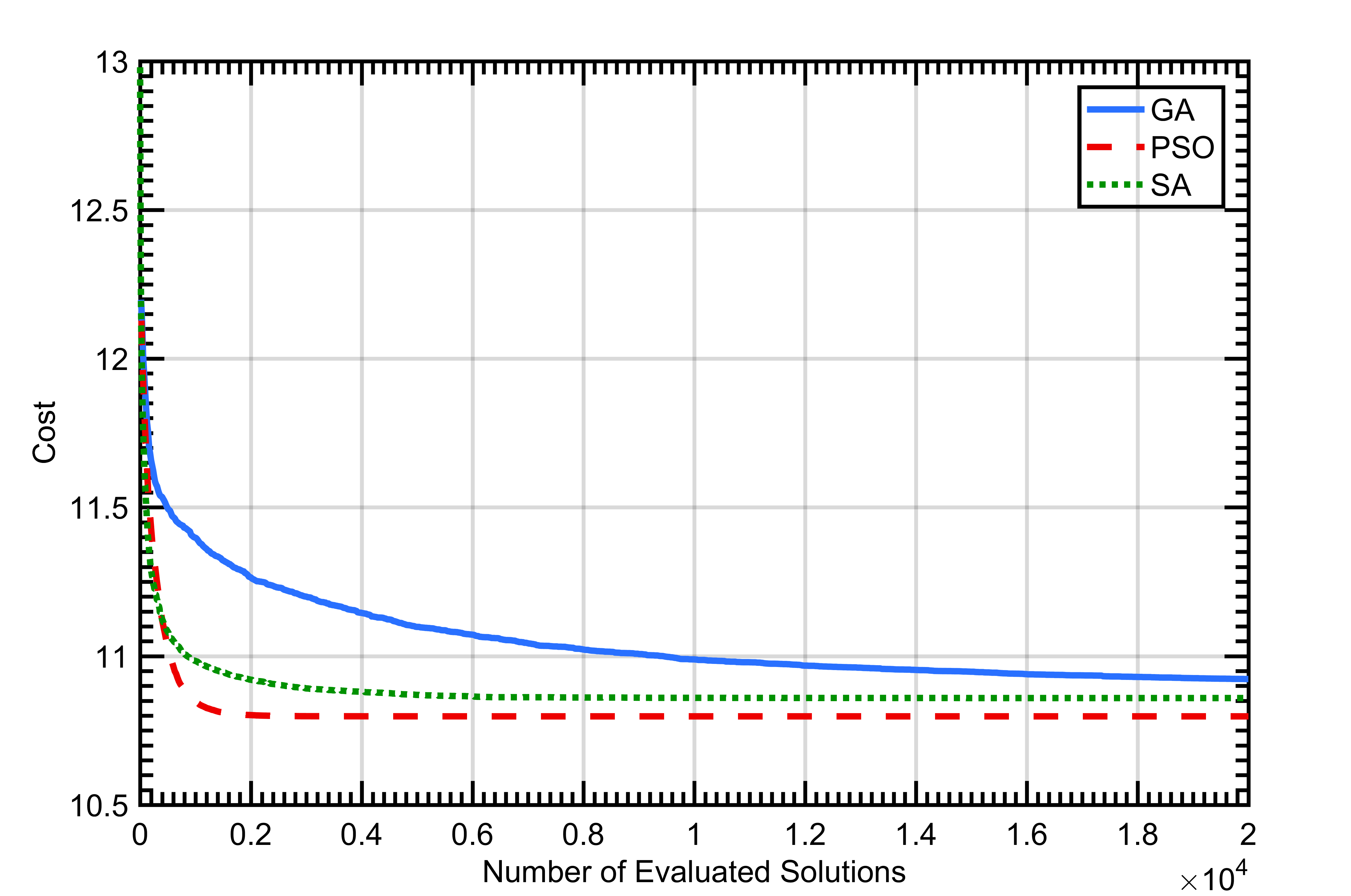}
            \caption{The mean value of the convergence curves (after 100 executions) for the three algorithms using the 40x40 array.}
            \label{fig:40x40_meancurves}
        \end{figure}

\begin{figure}[!ht]
            \centering
				\subfloat[GA\label{40x40_iterations_a}]{
						\includegraphics[width=1.65 in]{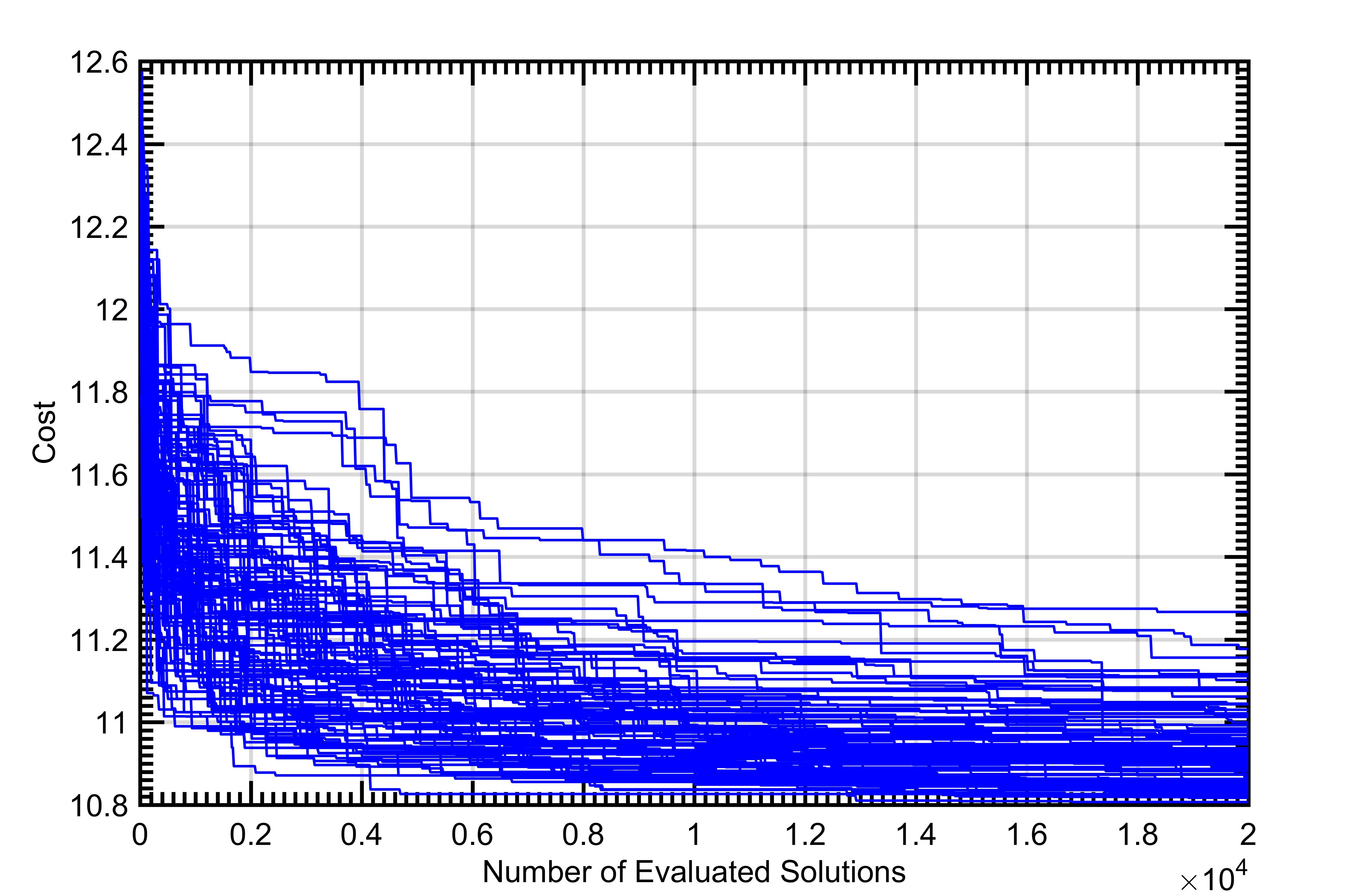}}
				\hfill
				\subfloat[PSO\label{40x40_iterations_b}]{
						\includegraphics[width=1.65 in]{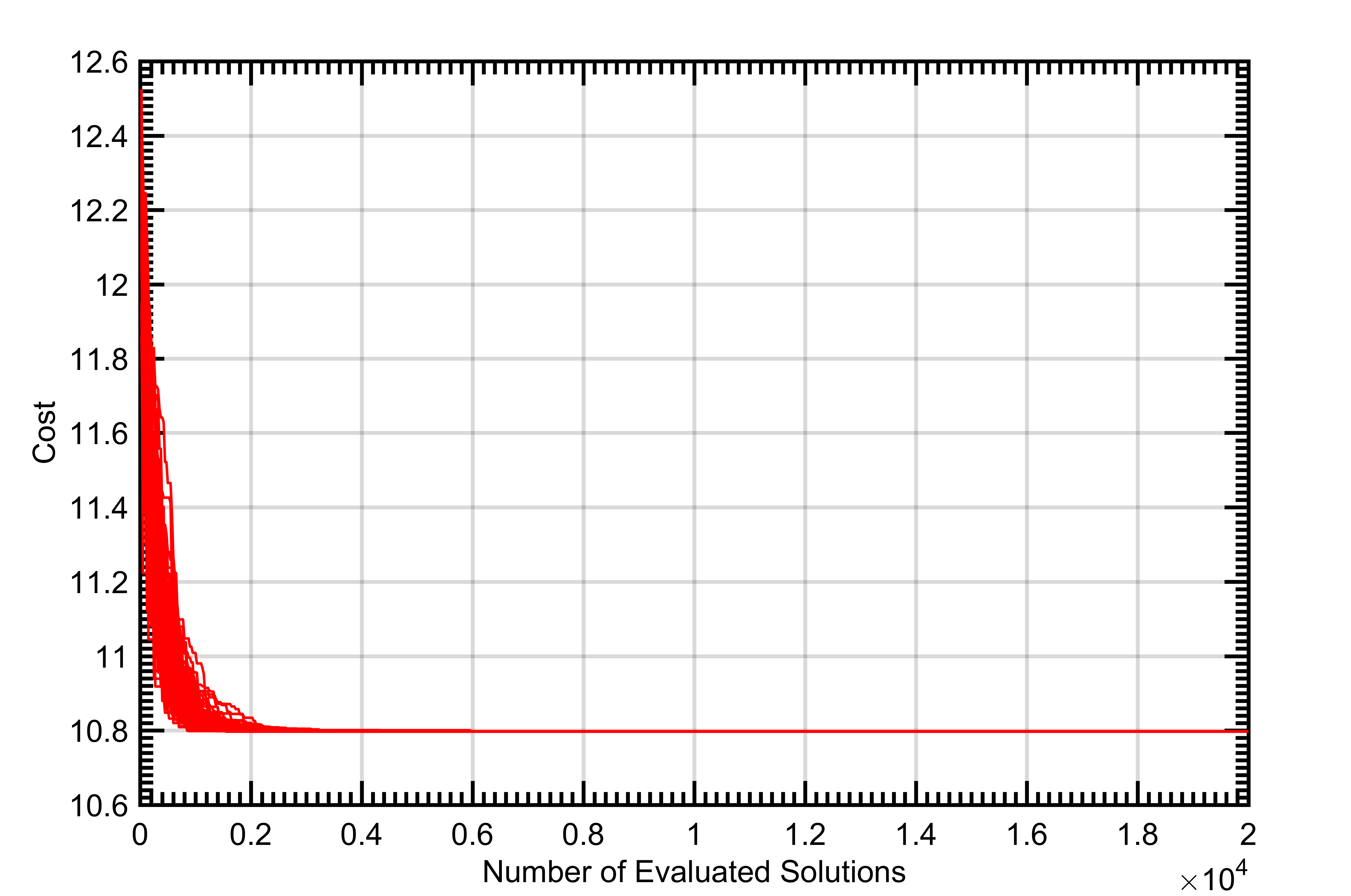}}
				\hfill
				\subfloat[SA\label{40x40_iterations_c}]{
						\includegraphics[width=1.65 in]{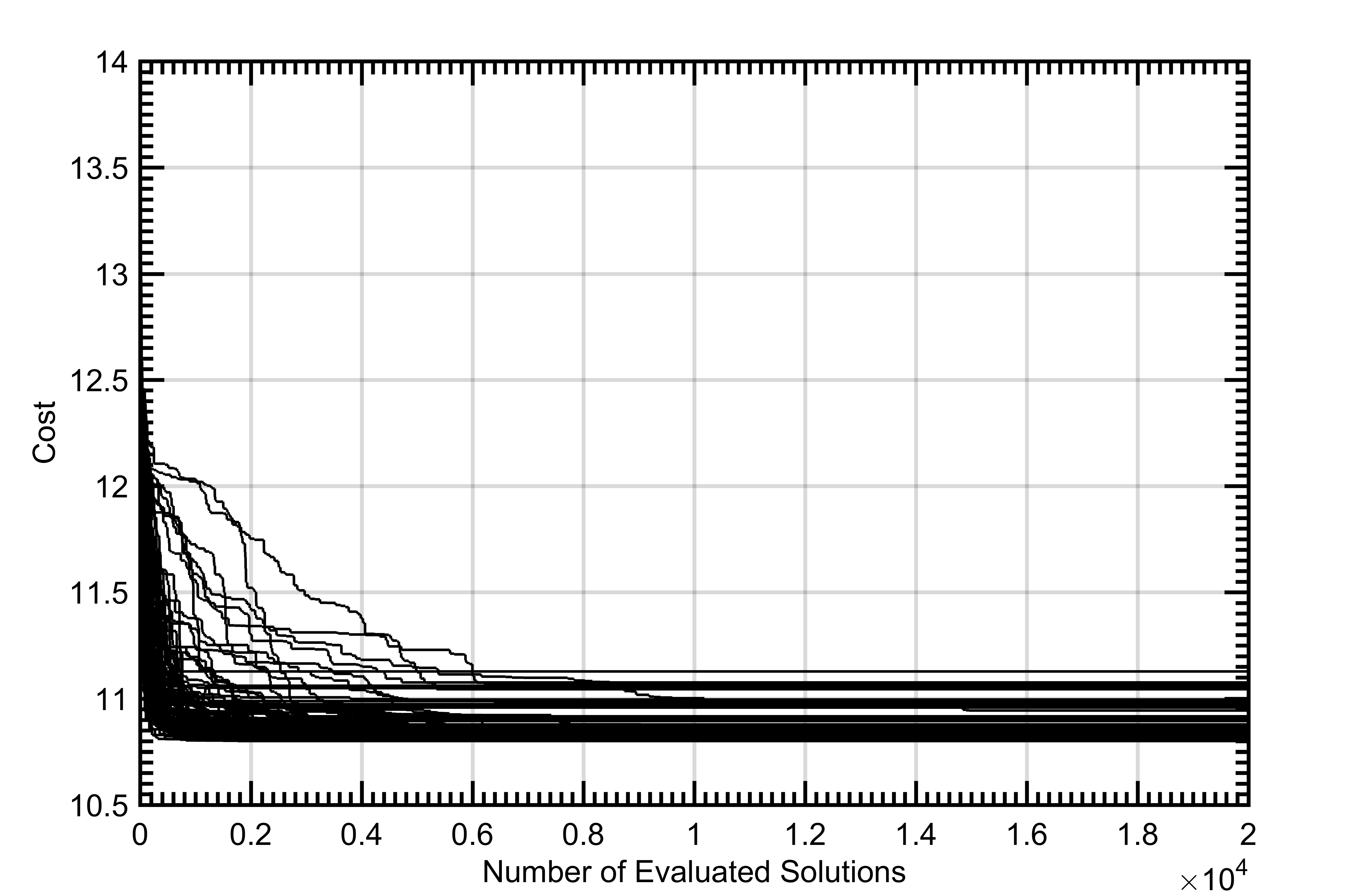}}
				\hfill
            \caption{The convergence curves for the synthesized 40x40 array for each of the three metaheuristic global optimization algorithm. Notice that each algorithm reaches the same best solution.}
            \label{fig:40x40_cost_executions}
        \end{figure}

        All the algorithms performed well on the 40x40 array. As can be seen in Figure \ref{fig:40x40_meancurves}.
        On average, PSO and SA performed more efficiently than GA. However, the best solutions from the repeated iterations of all the algorithms resulted in the same cost solution that had the same phase values. Therefore, all three algorithms were equally effective and converged to the same global optimum solution for at least some of the 100 executions performed for each algorithm. Figure \ref{fig:40x40_cost_executions} shows the convergence of each of the 100 executions of each of the algorithms for the 40x40 arrays.
            
\begin{figure}[!ht]
        \subfloat[2D Pattern\label{40x40_pattern_a}]{
            \centering
                \includegraphics[width=3.2 in]{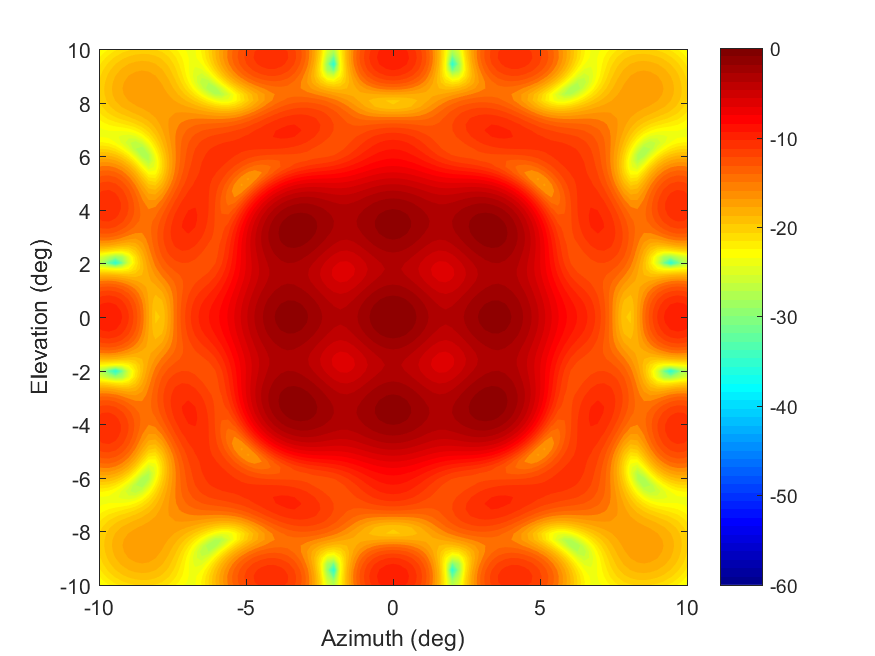} }
        
        \subfloat[1D Cuts of the 2D Pattern\label{40x40_cut}]{
            \centering
                \includegraphics[width=3.2 in]{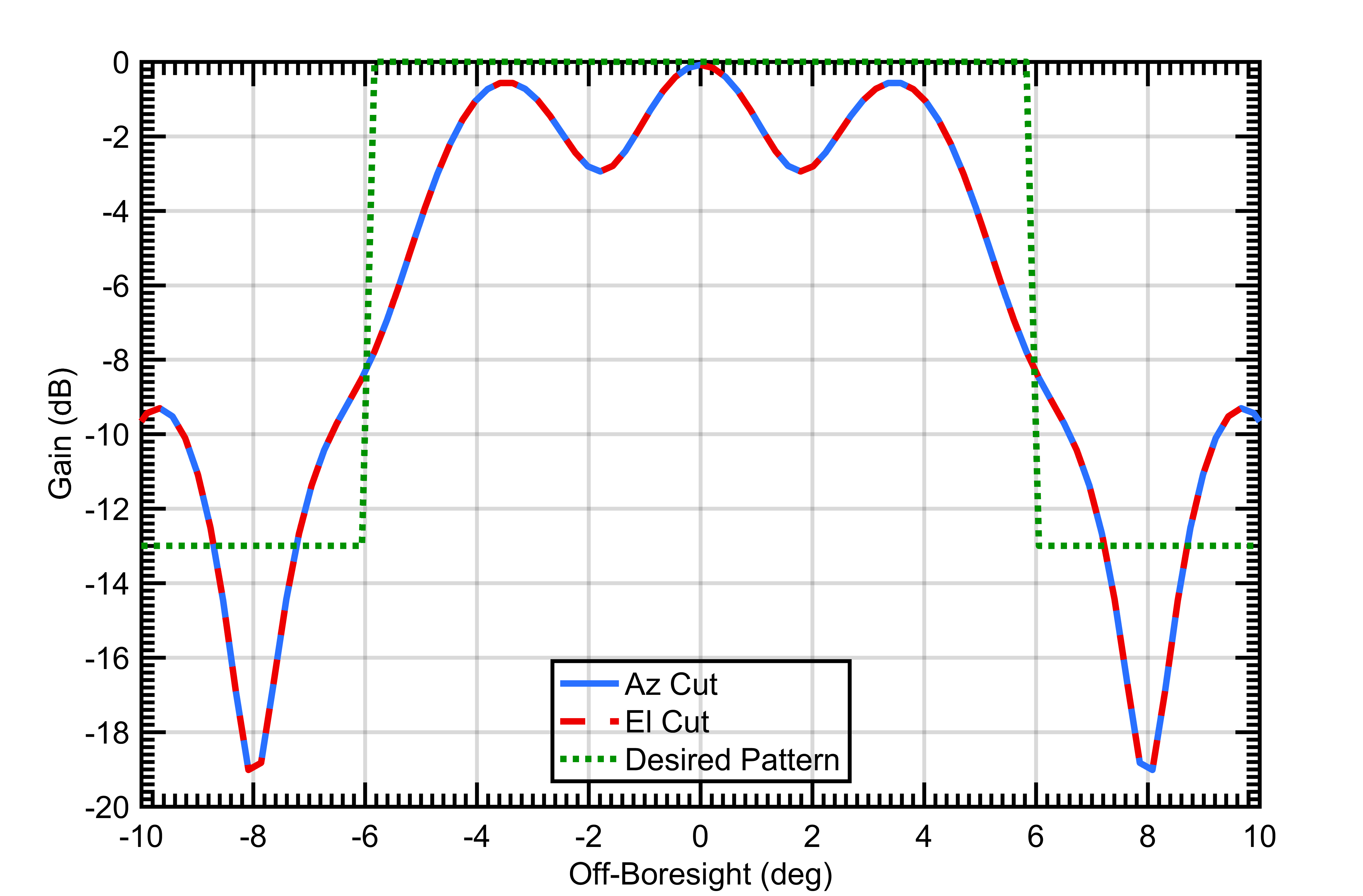} }
            \caption{Best resulting 40x40 array pattern (cost = 10.8) achieved by all three algorithms in at least some of the 100 executions of 20,000 evaluations. 2D cuts along the azimuth and elevation axes are shown. The array pattern at angles in the visible region not shown in this figure are below the -13 dB sidelobe level.} 
            \label{fig:40x40_pattern}
        \end{figure}

        The pattern of the best solution is seen in Figure \ref{fig:40x40_pattern}. The cost of the solution was 10.8 after the 20,000 evaluations of $C$.
        
\begin{figure}[!ht]
            \centering
                \includegraphics[width=3.2 in]{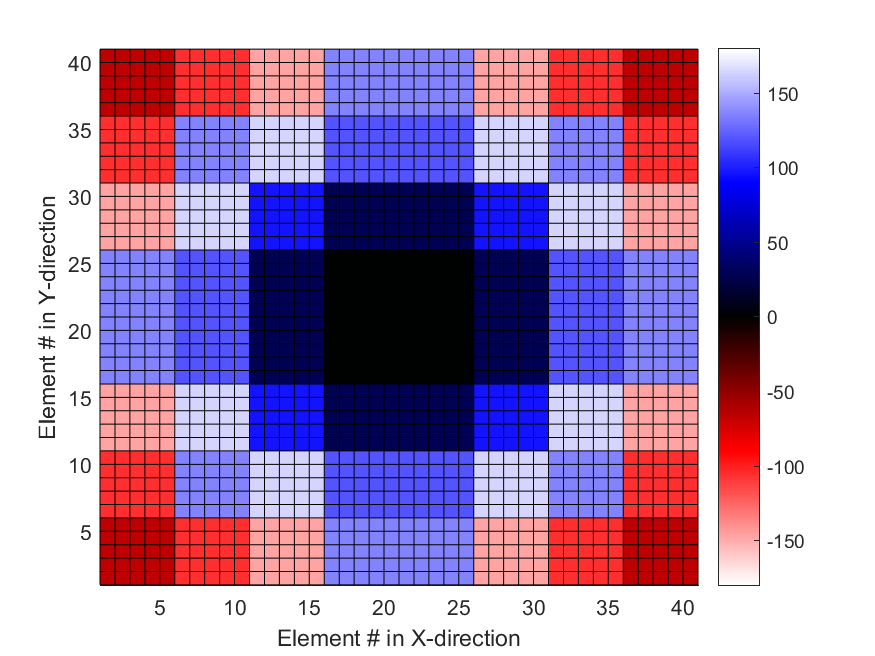}
            \caption{Phase values of the elements for the array that generated the pattern in Figure \ref{fig:40x40_pattern}.}
            \label{fig:40x40_phases}
        \end{figure}

        The phase values of the subarrays for this best solution are shown in Figure \ref{fig:40x40_phases}. 
        The phase values form somewhat concentric rings of similar phase values. This is difficult to visualize in this smaller-sized array and can be more clearly observed in the larger 80x80 arrays. 
        
    \subsection{80x80 Results}
    
\begin{figure}[!ht]
            \centering
				\subfloat[GA\label{80x80_iterations_a}]{
						\includegraphics[width=1.65 in]{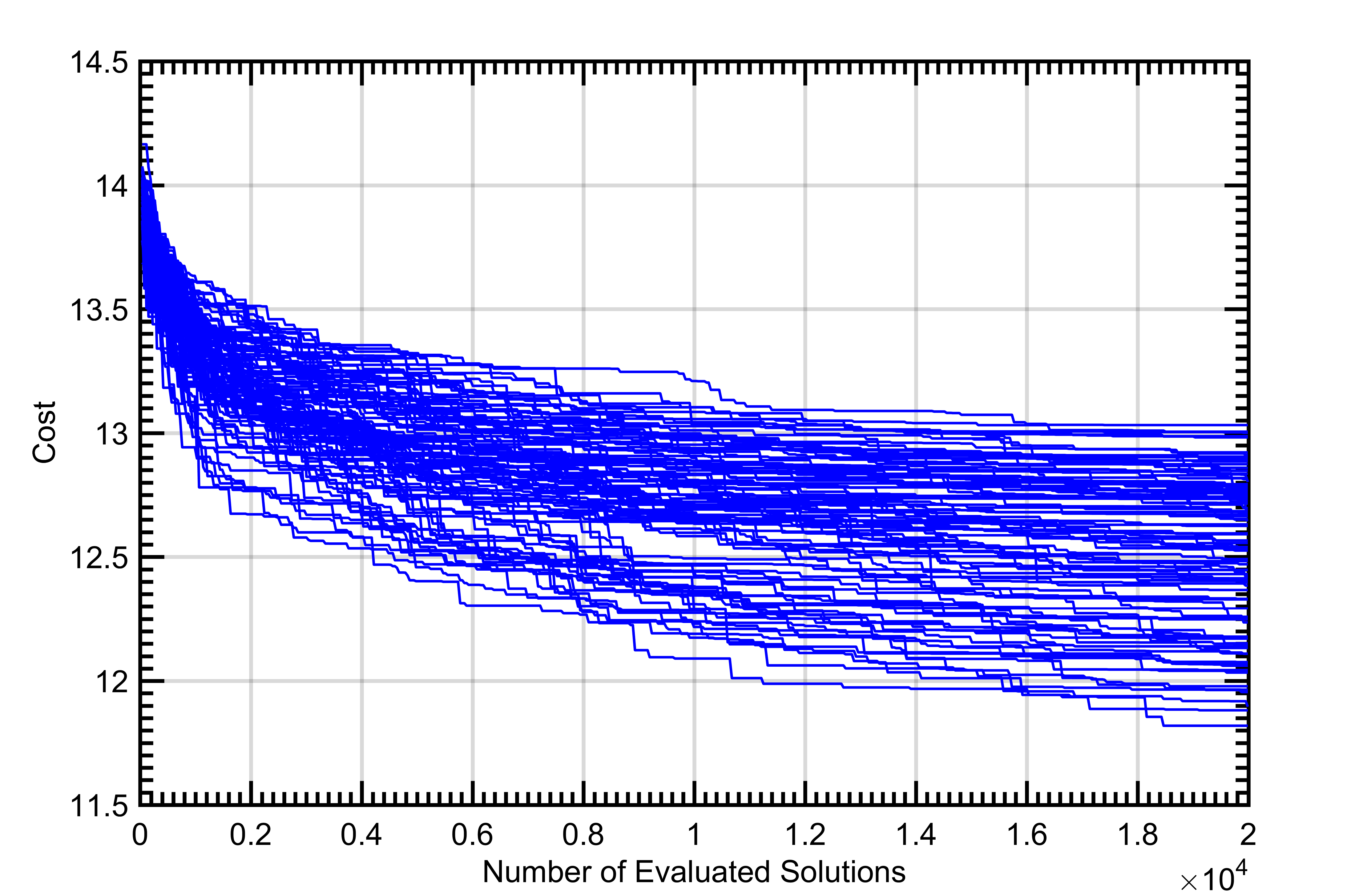}}
				\hfill
				\subfloat[PSO\label{80x80_iterations_b}]{
						\includegraphics[width=1.65 in]{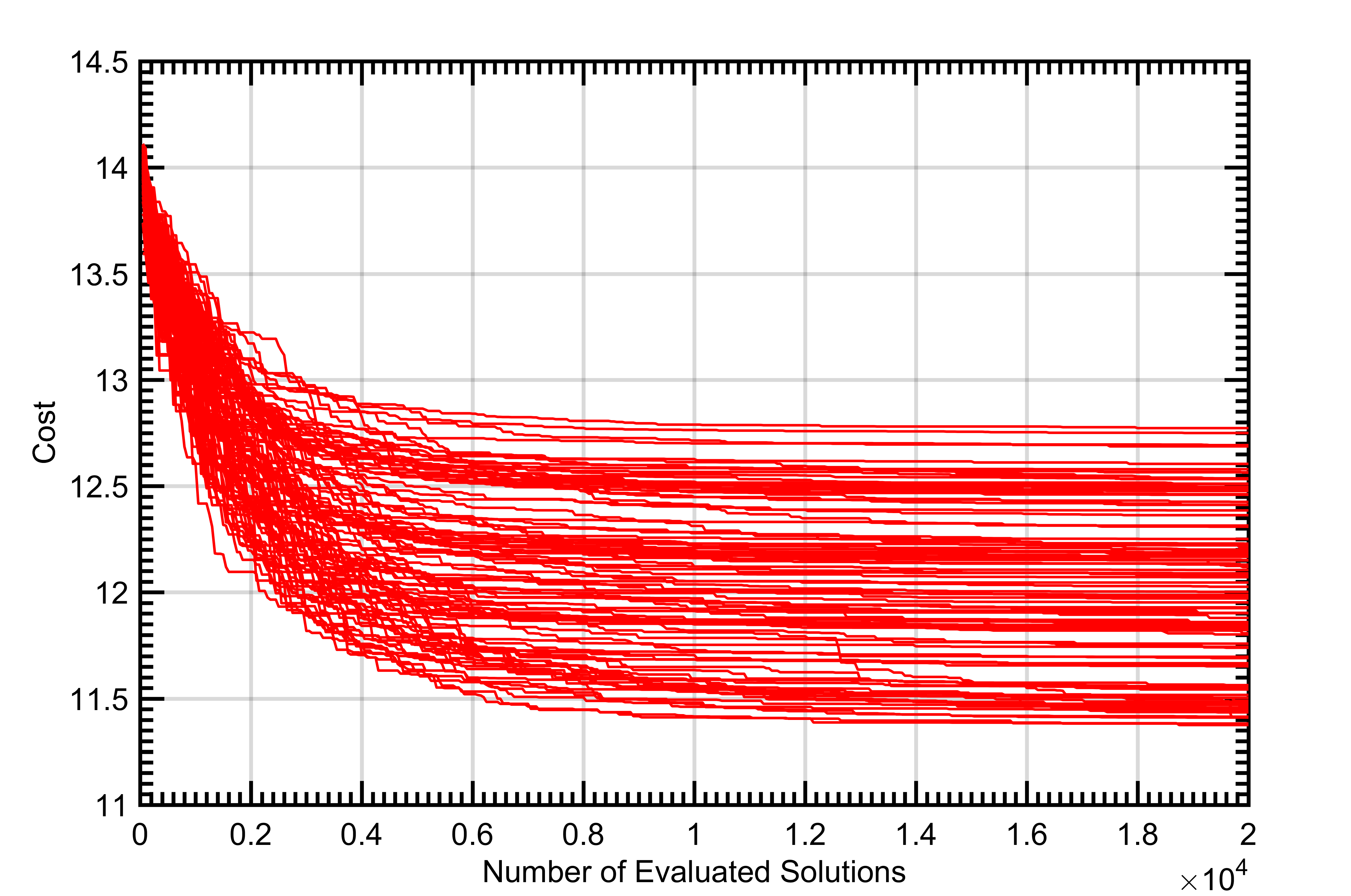}}
				\hfill
				\subfloat[SA\label{80x80_iterations_c}]{
						\includegraphics[width=1.65 in]{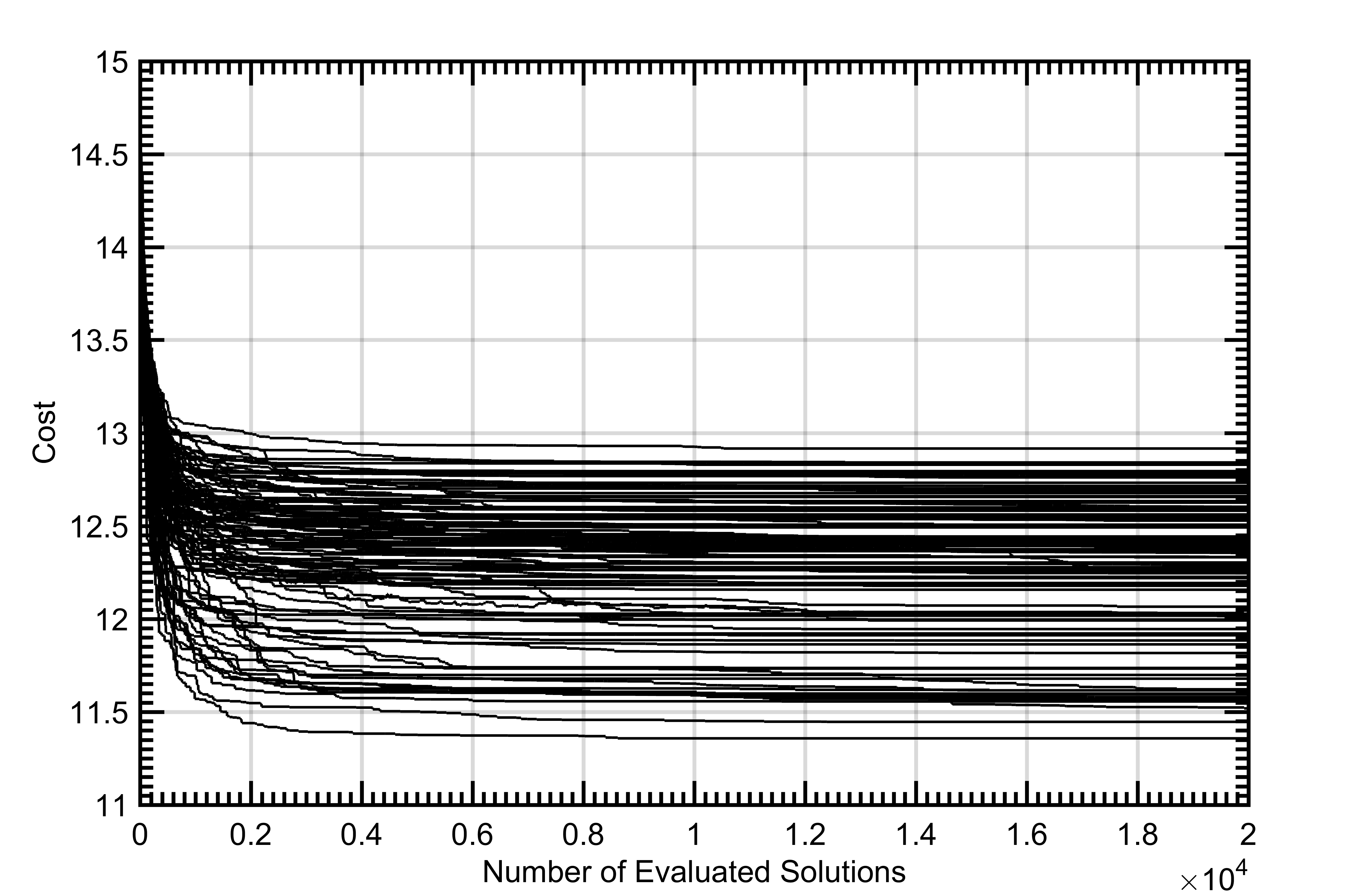}}
				\hfill
            \caption{The convergence curves for the synthesized 80x80 array for each of the three metaheuristic global optimization algorithm. Notice that the SA algorithm reaches the best solution but also has more problems getting stuck in local optima.  With the current parameter settings the SA algorithm rarely continues to converge with any significance after 5,000 iterations}
            \label{fig:80x80_cost_executions}
        \end{figure}

\begin{figure}[!ht]
            \centering
                \includegraphics[width=3.2 in]{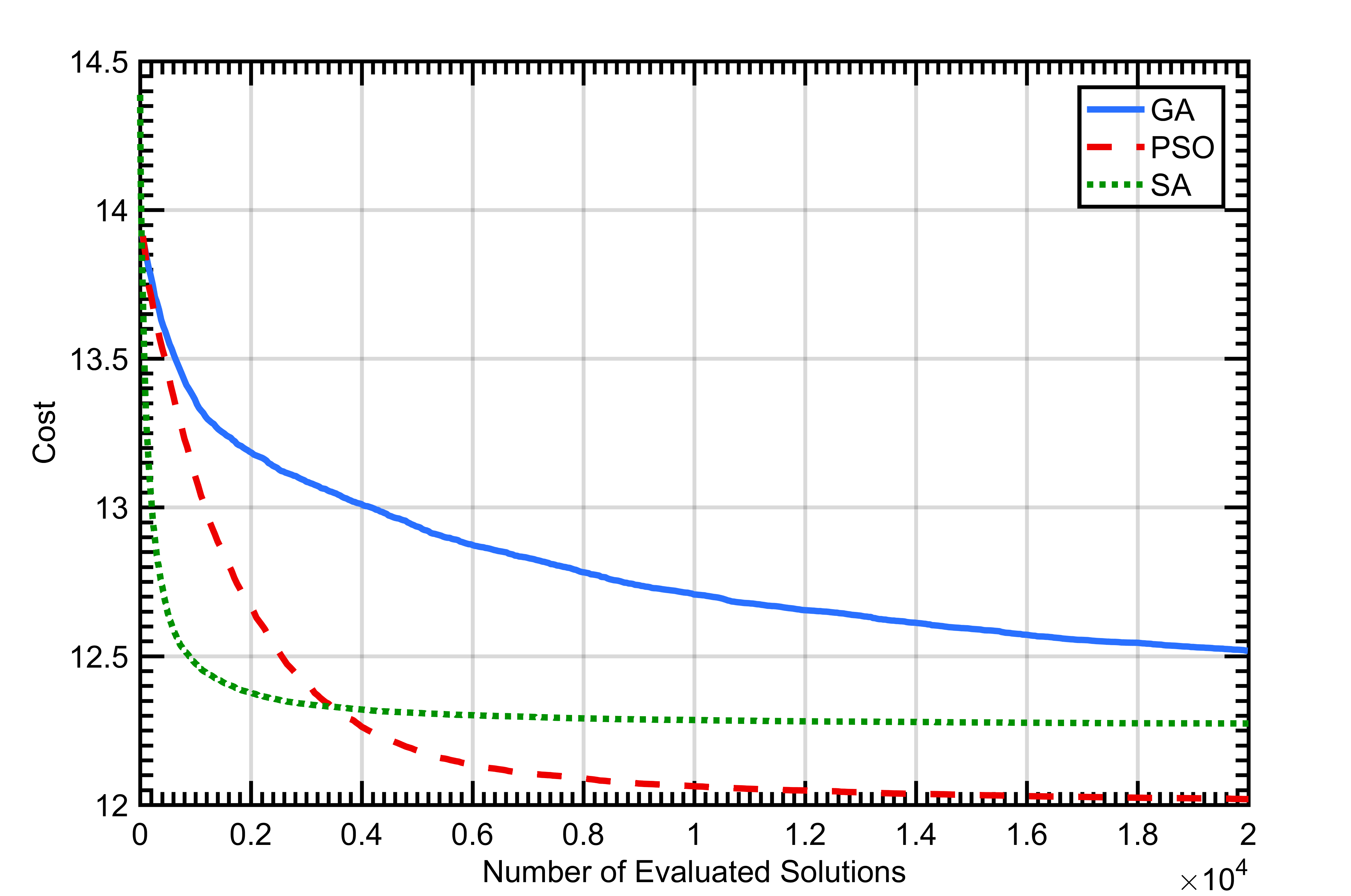}
            \caption{The mean value of the convergence curves for the three algorithms using the 80x80 array.}
            \label{fig:80x80_meancurves}
        \end{figure}

        The performance of the algorithms on the 80x80 array, as seen in Figure \ref{fig:80x80_meancurves}, varies more significantly than on the 40x40 array. In general, all three algorithms had a more difficult time finding the global optimum and were plagued with getting stuck in local optima. This struggle is shown in Figure \ref{fig:80x80_cost_executions} which shows the convergence of each of the 100 executions of each of the algorithms for the 80x80 arrays.
        
\begin{figure}[!ht]
        
        \subfloat[2D Pattern\label{80x80_pattern_a}]{
            \centering
                \includegraphics[width=3.2 in]{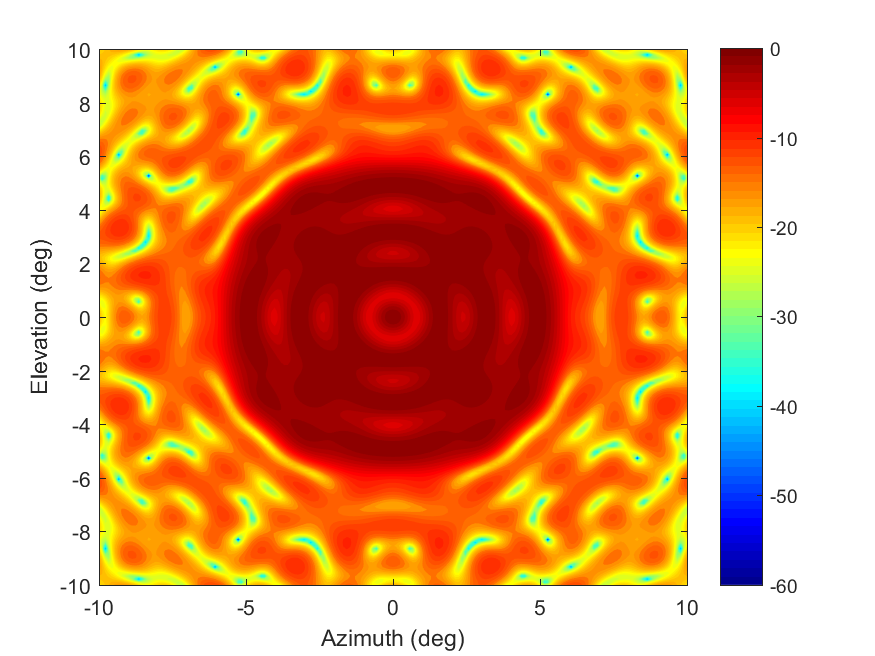}}
       
        \subfloat[1D Cuts of the 2D Pattern\label{80x80_cut}]{
            \centering 
                \includegraphics[width=3.2 in]{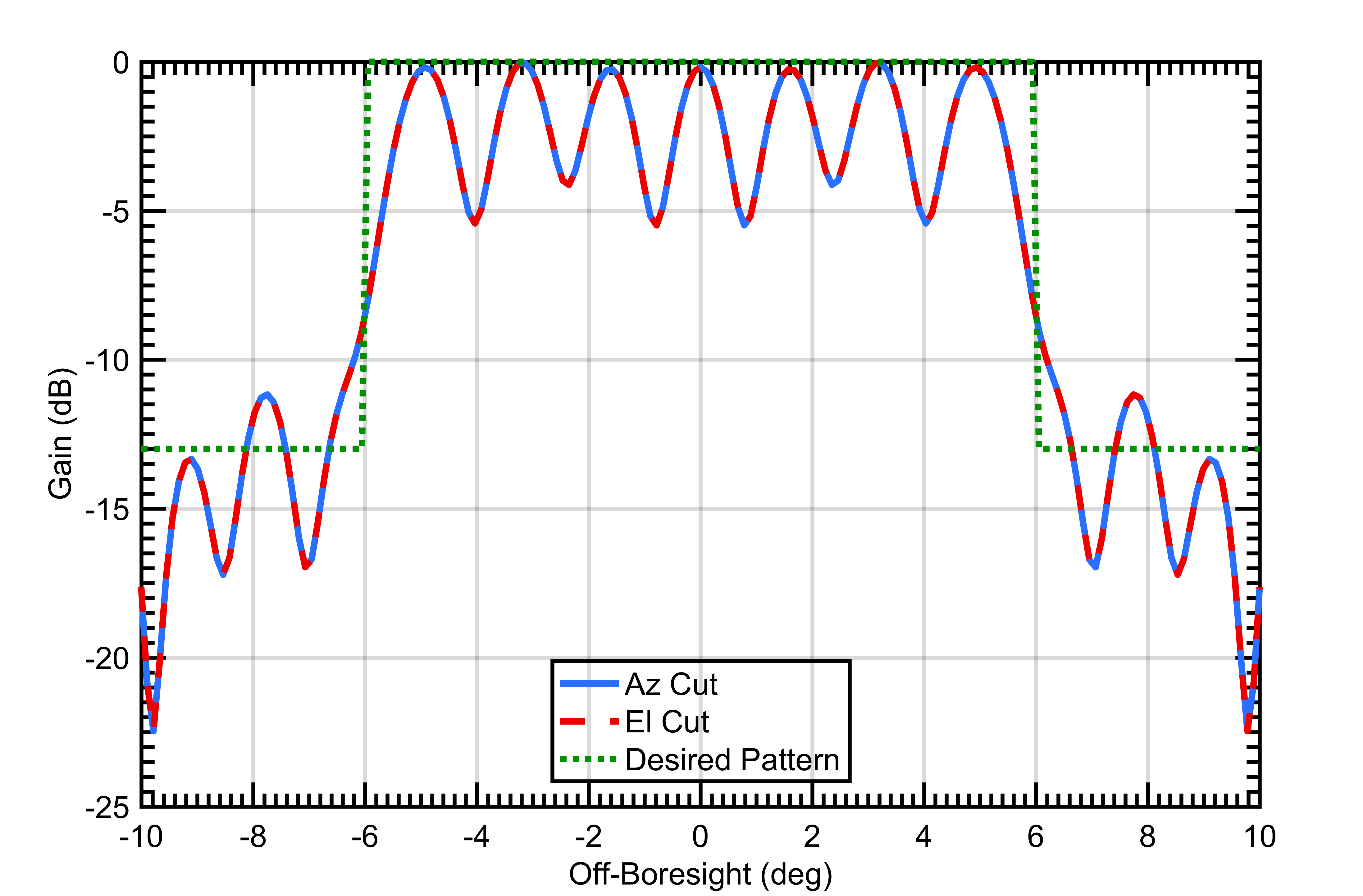} }
            \caption{Best resulting 80x80 array pattern (cost = 11.36) achieved by the SA algorithm in at least one of the 100 executions of 20,000 evaluations. 2D cuts along the azimuth and elevation axes are shown. The array pattern at angles in the visible region not shown in this figure are below the -13 dB sidelobe level.} 
            \label{fig:80x80_pattern}
        \end{figure}

        The pattern of the best solution for the 80x80 array is seen in Figure \ref{fig:80x80_pattern}. The cost of the best solution was 11.36 after the 20,000 evaluations of $C$.
        
\begin{figure}[!ht]
            \centering
                \includegraphics[width=3.2 in]{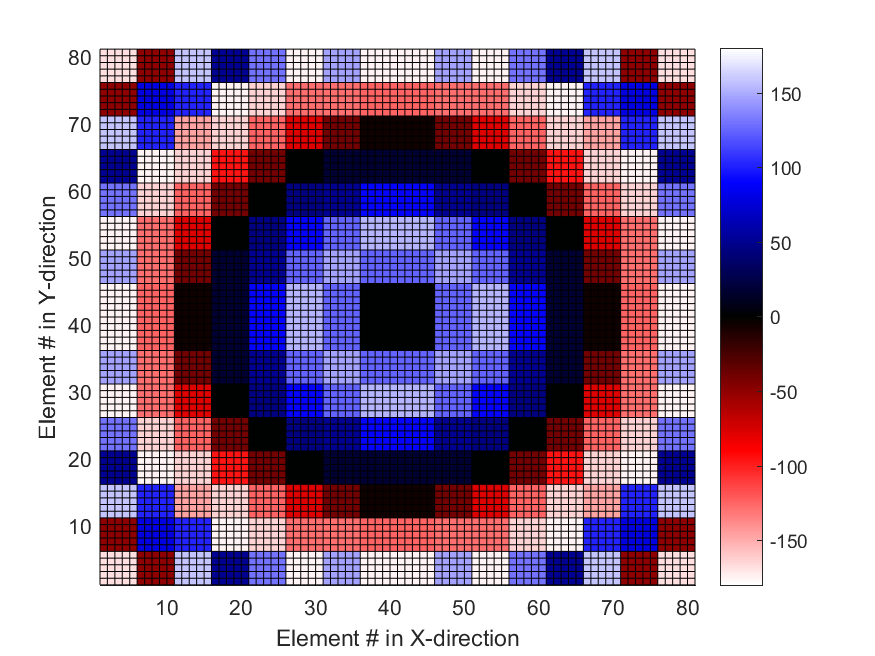}
            \caption{Phase values of the elements for the array that generated the pattern in Figure \ref{fig:80x80_pattern}.}
            \label{fig:80x80_phases}
        \end{figure}

        The phase values of the subarrays for this best solution are shown in Figure \ref{fig:80x80_phases}. Concentric rings of phase are clearly visible. 
    
\section{Summary and Conclusions}
\label{Conclusions Section}

    In the 40x40 radar architecture, all three algorithms in this study found the global optimum on at least some of their 100 executions.  On average, PSO was slightly more effective than SA. While SA started with the steepest convergence (Figure \ref{fig:40x40_meancurves}), it tended to get stuck in local optima more frequently than PSO (Figure \ref{fig:40x40_cost_executions}). Both PSO and SA outperformed genetic algorithm.  However, it should be noted that each metaheuristic technique can be configured to manage the speed of convergence and the probability that the convergence stalls in a local optima instead of the global optimum. The overall poorer performance of the GA when compared to the PSO and SA is likely influenced by the GA configuration and would likely have more comparable performance if more time were spent optimizing the configuration. This increased optimization of the GA algorithm is outside the scope of this study and is an area of future work. 
    
    In the 80x80 radar architecture, the algorithms showed larger differences in their efficiency and effectiveness. Simulated annealing again showed the steepest convergence decent (Figure \ref{fig:80x80_meancurves}) and did find the best solution of any of the algorithms ($C = 11.36$) but still suffered from stalling in local optima (Figure \ref{fig:80x80_cost_executions}). This effect is realized by PSO showing better average effectiveness than SA even though SA's best solution was slightly better than PSO's best solution.  Again, GA performed worse than SA and PSO, but it had less difficulty with stalling in local optima. In Figure \ref{fig:80x80_pattern} it should be noted that the ripples in the array pattern are deeper on the azimuth and elevation axes than at other orientations. While the nulls in the mainbeam are deeper than desired, actual measurements of similar arrays show that measured results typically smooth the nulls in simulated patterns so that in reality the nulls are not as deep as simulation can sometimes predict.

	In this study, we explore the ability of SA, GA, and PSO to generate desired beam broadened array patterns in subarrayed arrays using phase-only modification of element excitations. We have adapted and configured each algorithm for use with the problem domain, especially the PSO algorithm where the velocity calculation needed to be modified because phase is a circular function. We find that SA and PSO are more effective and efficient than GA for the configurations shown in this study. SA is more efficient than PSO, but PSO is on average more effective than SA.

	Future studies can examine these techniques for larger subarrayed arrays to better understand their effectiveness. Comparisons of these techniques utilized for both subarrayed and non-subarrayed arrays, with and without symmetry, is another potential path for research. Also, utilizing a supercomputing cluster could allow verification of the global optimum for these array configurations and better quantify the optimality of solutions found from these techniques.

\bibliographystyle{IEEEtran}
\bibliography{BeamBroadeningGlobalOptimization}

\end{document}